\documentclass{jfm}

\usepackage{graphicx}
\usepackage{natbib}

\ifCUPmtlplainloaded \else
  \checkfont{eurm10}
  \iffontfound
    \IfFileExists{upmath.sty}
      {\typeout{^^JFound AMS Euler Roman fonts on the system,
                   using the 'upmath' package.^^J}%
       \usepackage{upmath}}
      {\typeout{^^JFound AMS Euler Roman fonts on the system, but you
                   dont seem to have the}%
       \typeout{'upmath' package installed. JFM.cls can take advantage
                 of these fonts,^^Jif you use 'upmath' package.^^J}%
      }
  \else
  \fi
\fi

\ifCUPmtlplainloaded \else
  \checkfont{msam10}
  \iffontfound
    \IfFileExists{amssymb.sty}
      {\typeout{^^JFound AMS Symbol fonts on the system, using the
                'amssymb' package.^^J}%
       \usepackage{amssymb}%

      }{}
  \fi
\fi

\ifCUPmtlplainloaded \else
  \IfFileExists{amsbsy.sty}
    {\typeout{^^JFound the 'amsbsy' package on the system, using it.^^J}%
     \usepackage{amsbsy}}
    {}
\fi

\newcommand\Pran{\mbox{\textit{Pr}}} % Prandtl number, cf TeX's \Pr product
\newsavebox{\astrutbox}
\sbox{\astrutbox}{\rule[-5pt]{0pt}{20pt}}

\newcommand\ttz{\ensuremath{\rightarrow 0}}
\newcommand\ttinf{\ensuremath{\rightarrow \infty}}

\newcommand\be{\begin{equation}}
\newcommand\ee{\end{equation}}
\newcommand\bes{\begin{equation*}}
\newcommand\ees{\end{equation*}}
\newcommand{\pdv}[2]{\frac{\partial#1}{\partial#2}}
\newcommand{\dv}[2]{\frac{d#1}{d#2}}

\title[Analytical shock solutions]{Analytical shock solutions at large and small Prandtl number}

\author[B. M. Johnson]%
{B. M. Johnson$^1$%
  \thanks{Email address for correspondence: johnson359@llnl.gov}}

\affiliation{$^1$Lawrence Livermore National Laboratory, Livermore,
CA 94550, USA}

\pubyear{2010}
\volume{650}
\pagerange{119--126}
\date{20 March 2013; revised 14 May 2013; accepted 16 May 2013}
\begin{document}

\maketitle

\begin{abstract}
Exact one-dimensional solutions to the equations of fluid dynamics are derived in the $\Pran \ttinf$ and $\Pran \ttz$ limits (where $\Pran$ is the Prandtl number). The solutions are analogous to the $\Pran = 3/4$ solution discovered by Becker and analytically capture the profile of shock fronts in ideal gases. The large-$\Pran$ solution is very similar to Becker's solution, differing only by a scale factor. The small-$\Pran$ solution is qualitatively different, with an embedded isothermal shock occurring above a critical Mach number. Solutions are derived for constant viscosity and conductivity as well as for the case in which conduction is provided by a radiation field. For a completely general density- and temperature-dependent viscosity and conductivity, the system of equations in all three limits can be reduced to quadrature. The maximum error in the analytical solutions when compared to a numerical integration of the finite-$\Pran$ equations is \textit{O}$\left(\Pran^{-1}\right)$ as $\Pran \ttinf$ and \textit{O}$\left(\Pran\right)$ as $\Pran \ttz$.
\end{abstract}

\section{Introduction}

Although the power of numerical techniques makes them indispensable for solving the equations of fluid dynamics, analytical solutions, while difficult to find, remain useful for several reasons. They build physical intuition, they can be quickly evaluated over a wide dynamic range, and they can be used to verify numerical algorithms. One such solution was discovered by \cite{Becker22} under the assumptions of a steady-state, one planar dimension, constant viscosity, an ideal gas equation of state, and a fluid Prandtl number of $3/4$. It consists of implicit, closed-form expressions for the fluid variables and analytically captures the behavior of shocks in ideal gases with $\Pran = 3/4$. \cite{Thomas44}, \cite{Morduchow49}, \cite{Hayes60} and \cite{Iannelli13} extended Becker's solution to non-constant viscosity and conductivity, for both a power-law variation with temperature and a Sutherland viscosity model \citep{White91}. This is a more realistic assumption for gases, whose viscosity typically varies with temperature \citep{White91}. Approximate solutions for $\Pran \neq 3/4$ have also been derived by \cite{Khidr85}.

It is shown here that analogous solutions can be derived in both the $\Pran \ttinf$ and $\Pran \ttz$ limits, for both constant and non-constant viscosity and thermal conductivity. The transport properties of large Prandtl number fluids are dominated by momentum diffusion, whereas those of small Prandtl number fluids are dominated by thermal diffusion. Becker's solution applies to air and many other gases, which have $\Pran \sim 0.75$. Examples at the other extremes include mercury ($\Pran \sim 10^{-2}$), gas mixtures ($\Pran \sim 10^{-1}$), engine oil ($\Pran \sim 10^2-10^5$) and the Earth's mantle ($\Pran > 10^{23}$) \citep{Clay73,Kaminski03,Bejan04,Hogg12}. In addition, plasmas behave as small-$\Pran$ fluids due to the importance of heat conduction by both electrons and radiation \citep{Zel'dovich02}. A proton-electron plasma, for example, has $\Pran = 0.065$ \citep{Chapman39}. It should be noted that not all of these fluids obey an ideal gas equation of state and other physics may need to be taken into account; see the discussion in \S\ref{sec:discussion}. Taken together with Becker's solution, the solutions derived here yield analytical profiles of shock fronts in ideal gases over a wide range of parameter space. The basic equations are outlined in \S\ref{sec:equations}, \S\ref{sec:solutions} gives the derivation of the solutions, and \S\ref{sec:discussion} discusses some implications.
 
\section{Basic equations}\label{sec:equations}

For a fluid with mass density $\rho$, velocity magnitude $v$, pressure $p$, internal energy $e$, temperature $T$, viscosity $\mu$ (this can be regarded as either the dynamic viscosity in the limit of negligible bulk viscosity, or the sum of the dynamic viscosity and $3/4$ the bulk viscosity), and thermal conductivity $\kappa$ ($\Pran = \mu C_p/\kappa$, where $C_p$ is the specific heat at constant pressure), the equations of fluid dynamics in planar geometry are:
\be\label{eq:continuity}
\pdv{\rho}{t} + \pdv{}{x}\left(\rho v\right) = 0,
\ee
\be
\pdv{}{t}\left(\rho v\right) + \pdv{}{x}\left(\rho v^2 + p - \frac{4\mu}{3} \pdv{v}{x}\right) = 0, 
\ee
\be\label{eq:energy}
\pdv{}{t}\left(\frac{1}{2}\rho v^2 + \rho e\right) + \pdv{}{x}\left(\rho v\left[\frac{1}{2} v^2 + h\right] - \frac{4\mu}{3} v \pdv{v}{x} - \kappa \pdv{T}{x}\right) = 0,
\ee
where $h = e + p/\rho$ is the fluid enthalpy \citep{Landau87}. It will be assumed throughout that the fluid obeys an ideal gas equation of state:
$$
p = \left(\gamma - 1\right) \rho e,
$$
so that $h = \gamma e = C_p T$ with $C_p = \gamma C_v$, where $C_v$ is the specific heat at constant volume. Under this assumption and the assumption of a steady-state, equations (\ref{eq:continuity})--(\ref{eq:energy}) can be integrated from $-\infty$ to $x$ to give:
\be\label{eq:mass_flux}
\rho v = \rho_0 v_0 \equiv m_0,
\ee
\be\label{eq:momentum_flux}
v^2 + \frac{\gamma - 1}{\gamma} h -\frac{4\mu}{3m_0} v\dv{v}{x} = \left(v_0^2 + \frac{\gamma - 1}{\gamma} h_0\right)\frac{\rho_0}{\rho},
\ee
\be\label{eq:energy_flux}
\frac{1}{2}v^2 + h - \frac{4\mu }{3 \rho} \dv{v}{x} - \frac{\kappa}{\rho v C_p} \dv{h}{x} = \left(\frac{1}{2}v_0^2  + h_0\right)\frac{\rho_0 v_0}{\rho v},
\ee
where the zero-slope boundary conditions appropriate for a shock are assumed to hold at $x = \pm \infty$. A subscript ``0'' here denotes a fluid quantity in the ambient (pre-shock) state. These equations can be combined into two ordinary differential equations governing the spatial profile of the shock front: 
\be\label{eq:veq}
\frac{4\mu}{3m_0} v\dv{v}{x} = v^2 + \frac{\gamma - 1}{\gamma} h - \frac{\gamma+1}{2 \gamma}\left(v_0 + v_1\right) v,
\ee
\be\label{eq:heq}
\frac{\kappa}{m_0 C_p} \dv{h}{x} = \frac{h}{\gamma} - \frac{v^2}{2} + \frac{\gamma+1}{2 \gamma}\left(v_0 + v_1\right) v - \frac{\gamma+1}{\gamma - 1}\frac{v_0v_1}{2},
\ee
where the integration constants have been expressed in terms of both pre-shock and post-shock (denoted by a subscript ``1'') velocities via the Rankine-Hugoniot jump conditions:
\be\label{eq:v1}
\frac{v_1}{v_0} = \frac{\gamma - 1 + 2/M_0^2}{\gamma+1},
\ee
where $M_0^2 = v_0^2/c_0^2$ is the shock Mach number and $c_0 = \sqrt{\gamma p_0/\rho_0}$ is the adiabatic sound speed in the ambient fluid \citep{Landau87}.

\section{Solutions}\label{sec:solutions}

The derivation of the \cite{Becker22} solution is outlined in \S\ref{sec:pran34}, followed by a derivation of the $\Pran \ttinf$ and $\Pran \ttz$ solutions in \S\ref{sec:praninf} and \S\ref{sec:pranzero}, respectively. These are all derived for constant viscosity and conductivity; \S\ref{sec:grey_diffusion} shows how the solutions can be extended to non-constant viscosity and conductivity, using radiation heat conduction as an example. General expressions for the shock profiles in all three $\Pran$ limits under the assumption of a viscosity and conductivity that vary as powers of the density and temperature are derived in \S\ref{sec:power_law}.

\subsection{Becker $\left(\Pran = 3/4\right)$ solution}\label{sec:pran34}

\cite{Becker22} noticed that for $\Pran = 3/4$ equation (\ref{eq:energy_flux}) for the energy flux,
\be\label{eq:ener_flux_pran34}
\frac{v^2}{2}  + h - \frac{\kappa}{m_0 C_p} \dv{}{x}\left(\frac{v^2}{2} + h\right) = \frac{v_0^2}{2}  + h_0,
\ee
is linear and has the finite solution
\be\label{eq:ener_flux_constant}
\frac{v^2}{2}  + h = \frac{v_0^2}{2}  + h_0 = \frac{\gamma+1}{\gamma - 1}\frac{v_0v_1}{2},
\ee
where the second equality follows from the Rankine-Hugoniot conditions. Solving this equation for $h$ and inserting it into equation (\ref{eq:momentum_flux}) for the momentum flux leads to
\be\label{eq:veq_pran34}
v L_\kappa \frac{\kappa}{\kappa_0} \dv{v}{x} = \frac{\gamma + 1}{2} \left(v - v_0\right)\left(v - v_1\right),
\ee
where
$$
L_\kappa \equiv \frac{\kappa_0}{m_0 C_v}.
$$
Equation (\ref{eq:veq_pran34}) can be rewritten as an integral over the velocity,
\be\label{eq:integral_pran34}
x  = \frac{2L_\kappa }{\gamma + 1} \int \frac{\left(\kappa/\kappa_0\right)v}{\left(v - v_0\right)\left(v - v_1\right)} \, dv.
\ee
For constant $\kappa = \kappa_0$, this integral is given by (to within an arbitrary constant)
\be\label{eq:solution_pran34}
x = \frac{2L_{\kappa}}{\gamma + 1} \ln \left[\left(v_0 - v\right)^\frac{v_0}{v_0 - v_1}\left(v - v_1\right)^{-\frac{v_1}{v_0 - v_1}}\right].
\ee
Physical notation has been retained here as an aid to intuition; notice that $x = \pm \infty$ at $v = v_1$ and $v = v_0$, respectively. Defining the origin at the adiabatic sonic point $\left(v = \sqrt{v_0v_1}\right)$ and using $\eta \equiv v/v_0 = \rho_0/\rho$ (the specific volume relative to its ambient value) rather than $v$ yields the expression given in \cite{Zel'dovich02}. From (\ref{eq:ener_flux_constant}), the temperature in this limit is given by
\be\label{eq:temp_pran34}
T = \frac{R_\infty v_0 v_1 - v^2}{2 C_p},
\ee
where
$$
R_\infty \equiv \frac{\gamma+1}{\gamma - 1}
$$
is the maximum compression ratio.

\begin{figure}
\centering
\begin{tabular}{cc}
  \includegraphics[scale=0.4]{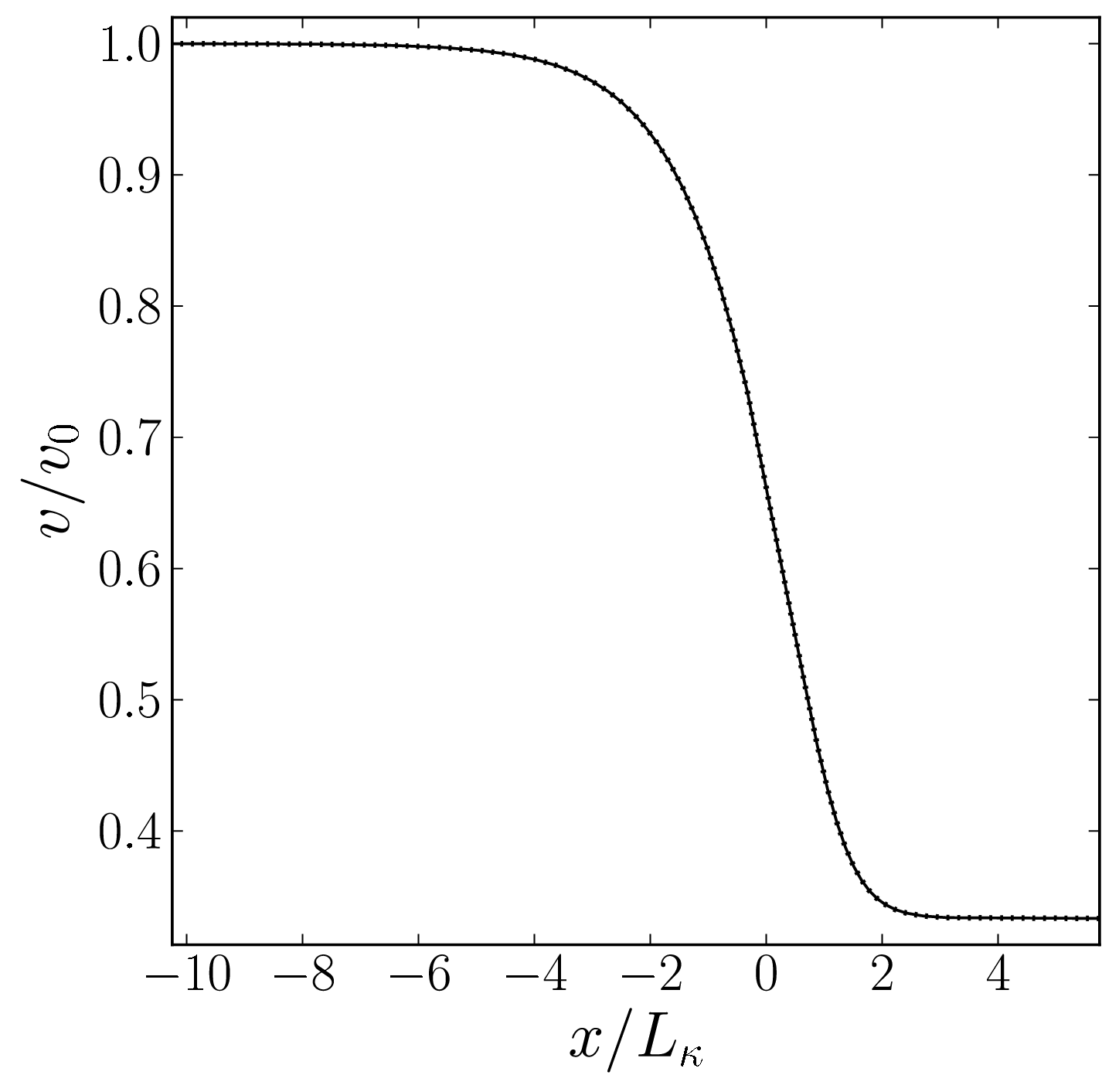} &
  \hspace{-0.0in}
  \includegraphics[scale=0.4]{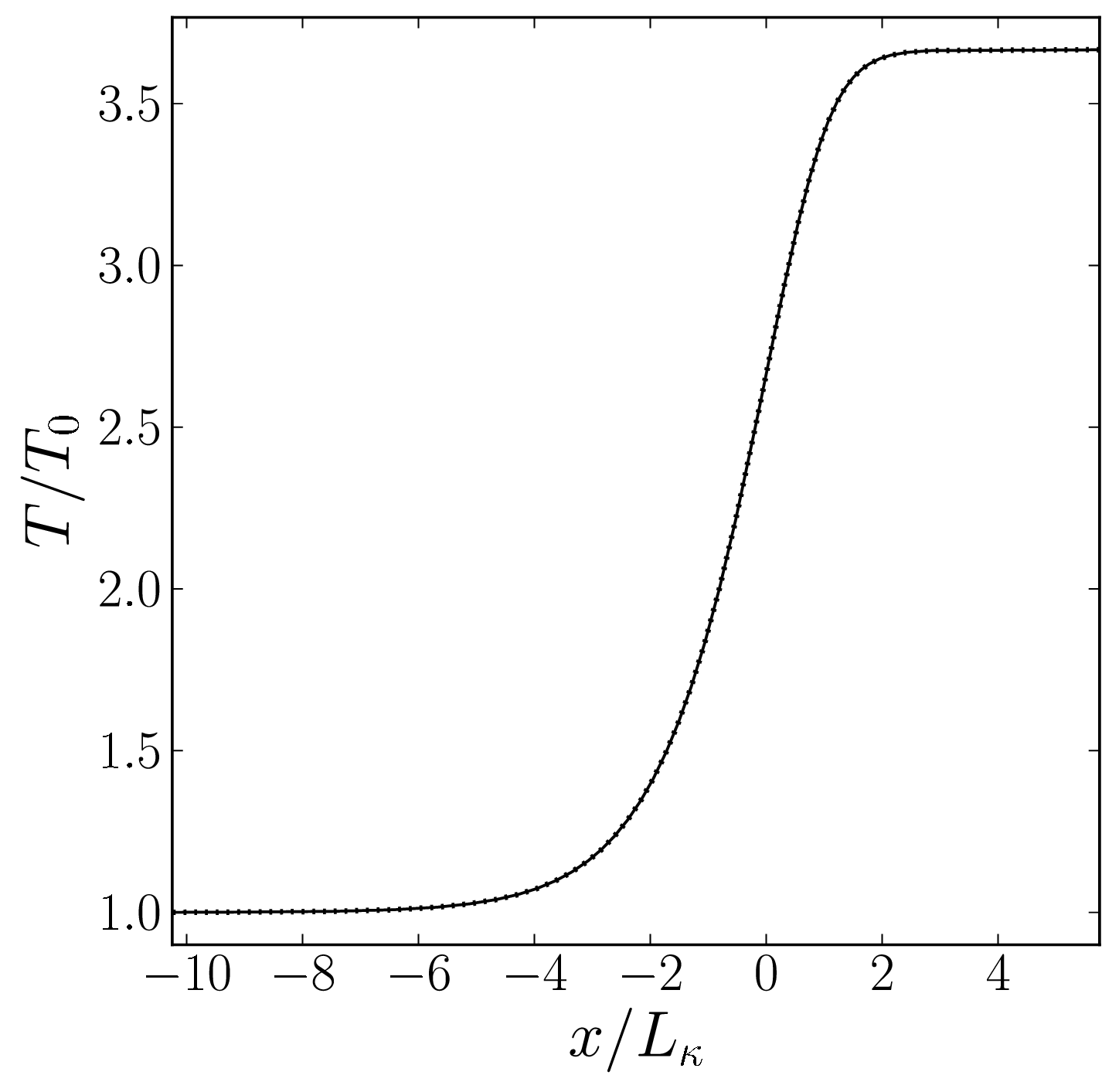}
\end{tabular}
  \caption{Velocity (\emph{left}) and temperature (\emph{right}) for the Becker solution ($\Pran = 3/4$) with $M_0 = 3$. No distinction is visible between the analytical (\emph{solid}) and numerical (\emph{dotted}) results.}
\label{fig:solution_pran34}
\end{figure}

Figure \ref{fig:solution_pran34} shows the velocity and temperature for this solution, using expressions (\ref{eq:solution_pran34}) and (\ref{eq:temp_pran34}). For comparison, results from a numerical integration of equations (\ref{eq:veq}) and (\ref{eq:heq}) are shown in figure \ref{fig:solution_pran34} as well. The numerical results here and in the following sections were obtained via a shooting method using the \texttt{odeint} differential equation solver in \texttt{scipy}. An important practical note here is that it is necessary to shoot from the post-shock state in order to obtain the desired solution. Equation (\ref{eq:ener_flux_pran34}) admits an exponential solution in addition to the constant solution, representing an additional energy flux at the boundary of arbitrary magnitude \citep{Zel'dovich02}. For an integration from the pre- to post-shock state, this solution is exponentially growing, bounded only by the end point of the integration, and can quickly dominate the numerical results. For an integration from the post- to pre-shock state, the exponential solution is decaying and therefore unproblematic.

\subsection{Large-$\Pran$ solution}\label{sec:praninf}

\begin{figure}
\centering
\begin{tabular}{cc}
  \includegraphics[scale=0.4]{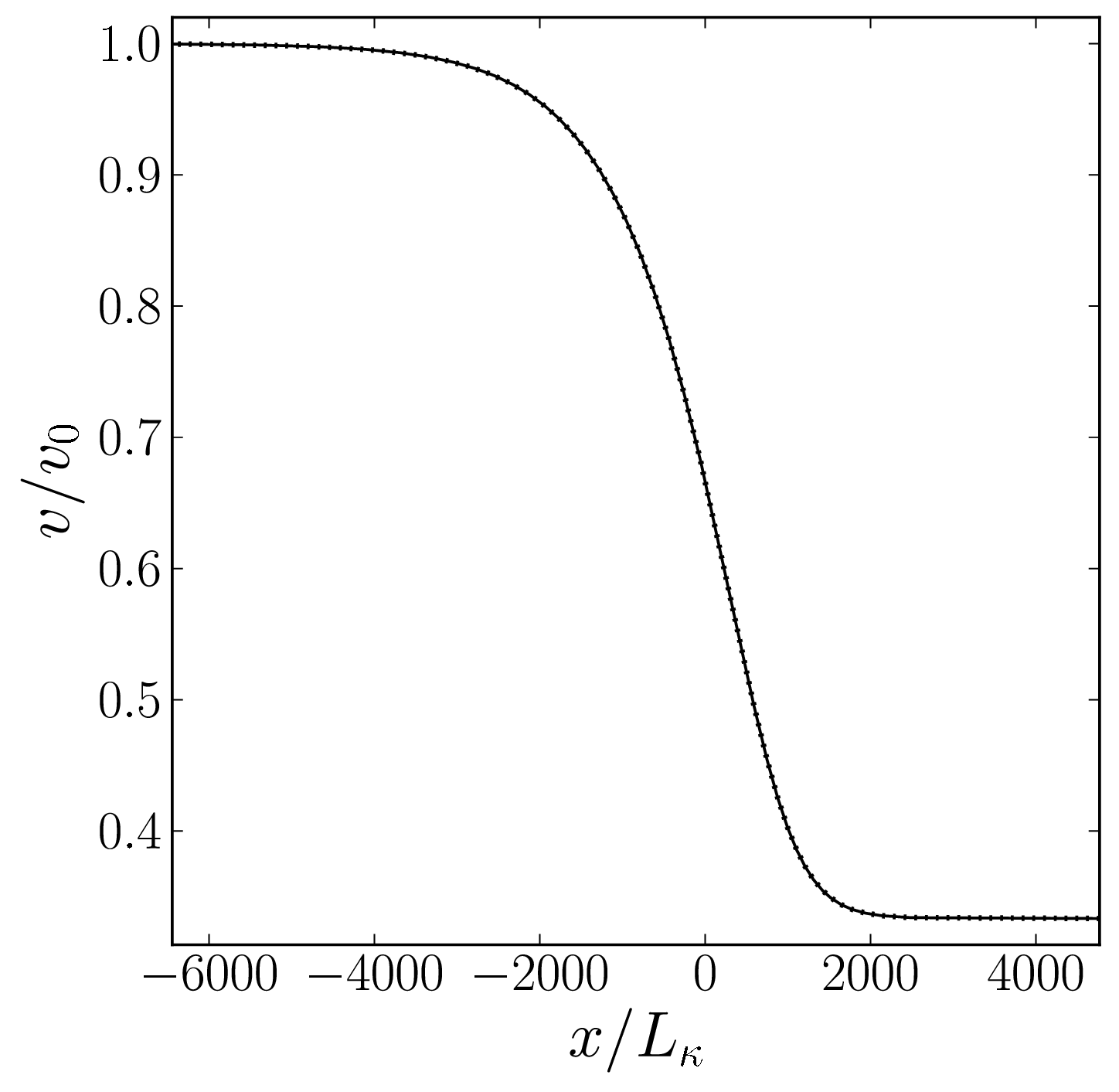} &
  \hspace{-0.0in}
  \includegraphics[scale=0.4]{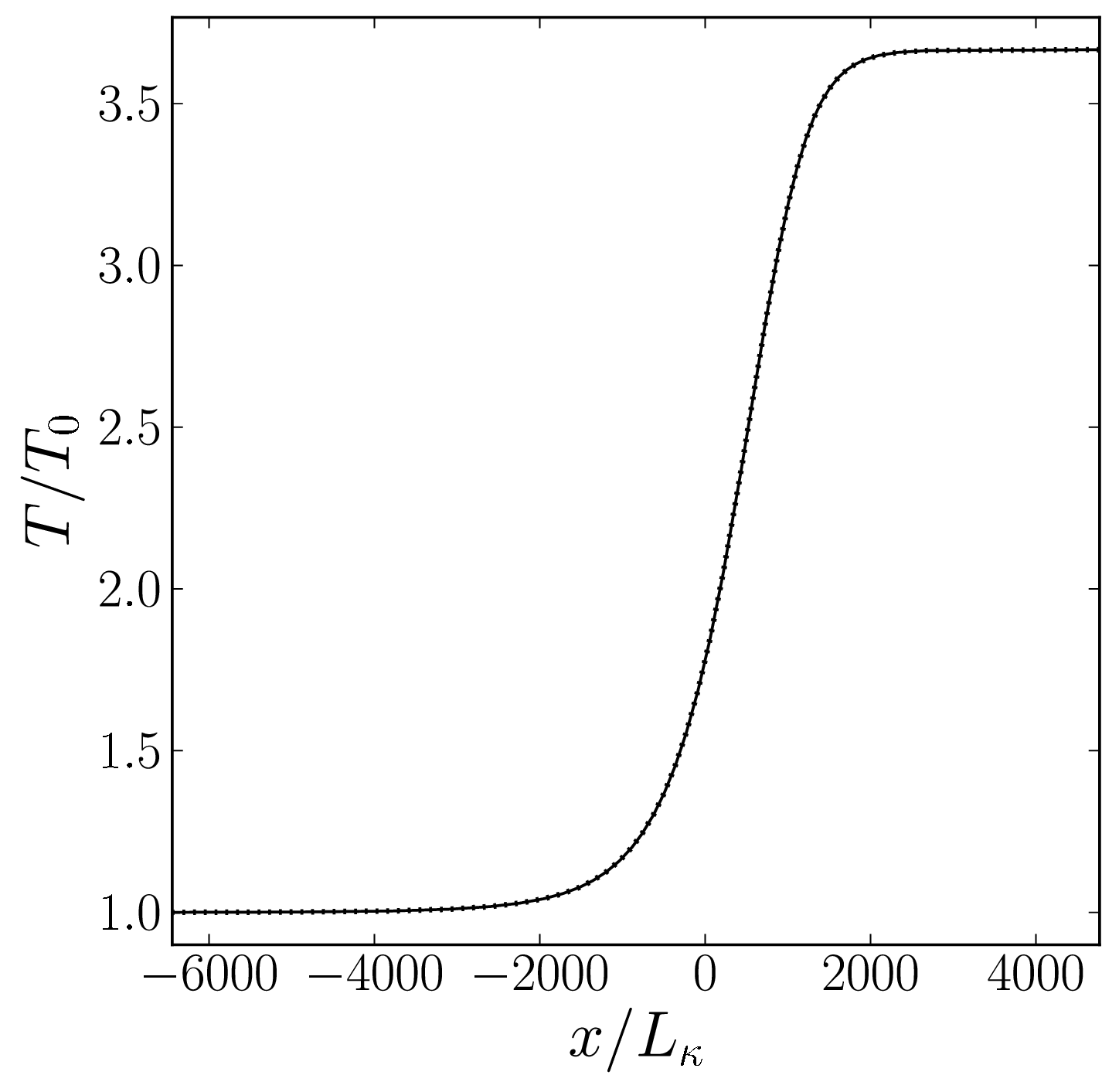}
\end{tabular}
  \caption{Velocity (\emph{left}) and temperature (\emph{right}) for the $\Pran \ttinf$ solution with $M_0 = 3$ and constant viscosity. No distinction is visible between the analytical (\emph{solid}) and numerical (\emph{dotted}) results.}
\label{fig:solution_praninf}
\end{figure}

In the limit $\Pran \ttinf$ ($\kappa \rightarrow 0$), equations (\ref{eq:momentum_flux}) and (\ref{eq:energy_flux}) become
\be\label{eq:mom_flux_praninf}
v^2 + \frac{\gamma - 1}{\gamma}h -\frac{4\mu}{3m_0} v\dv{v}{x} = \frac{\gamma+1}{2\gamma}\left(v_0 + v_1\right) v,
\ee
\be\label{eq:ener_flux_praninf}
\frac{1}{2}v^2 + h - \frac{4\mu}{3 m_0} v \dv{v}{x} = \frac{\gamma+1}{\gamma - 1}\frac{v_0v_1}{2},
\ee
which can be combined to give
\be\label{eq:veq_praninf}
v L_\mu \frac{\mu}{\mu_0} \dv{v}{x} = \frac{\gamma + 1}{2}\left(v - v_0\right)\left(v - v_1\right),
\ee
where
$$
L_\mu \equiv \frac{4\mu_0} {3 m_0} = \frac{4\Pran} {3 \gamma}L_\kappa.
$$
This can again be expressed as an integral over velocity,
\be\label{eq:integral_praninf}
x  = \frac{2 L_\mu}{\gamma + 1} \int \frac{\left(\mu/\mu_0\right) v}{\left(v - v_0\right)\left(v - v_1\right)} \, dv,
\ee
with the solution (for constant $\mu = \mu_0$) given by
\be\label{eq:solution_praninf}
x = \frac{2L_{\mu}}{\gamma + 1} \ln \left[\left(v_0 - v\right)^\frac{v_0}{v_0 - v_1}\left(v - v_1\right)^{-\frac{v_1}{v_0 - v_1}}\right].
\ee
Comparing expression (\ref{eq:solution_praninf}) with (\ref{eq:solution_pran34}), it can be seen that the velocity profile in the large-$\Pran$ solution differs from that of the \cite{Becker22} solution only by the scale factor $L_\mu/L_\kappa = 4\Pran/(3\gamma)$ (assuming constant $\Pran$). The difference between the temperature profiles is more complicated, since solving equations (\ref{eq:mom_flux_praninf}) and (\ref{eq:ener_flux_praninf}) for the temperature in this limit yields an expression that differs from expression (\ref{eq:temp_pran34}):
\be\label{eq:temp_praninf}
T = \frac{v^2 - 4v_i v + R_\infty v_0 v_1}{2 C_v},
\ee
where
\be\label{eq:etai}
v_i \equiv \frac{\gamma+1}{4\gamma}\left(v_0 + v_1\right).
\ee
Figure \ref{fig:solution_praninf} shows the velocity and temperature for the large-$\Pran$ solution with $M_0 = 3$ and constant viscosity. A value of $\Pran = 10^{3}$ was used to generate the numerical results in this figure. 

\subsection{Small-$\Pran$ solution}\label{sec:pranzero}

In the limit $\Pran \ttz$ ($\mu \rightarrow 0$), equations (\ref{eq:momentum_flux}) and (\ref{eq:energy_flux}) become
\be\label{eq:mom_flux_pranzero}
v^2 + \frac{\gamma - 1}{\gamma} h = \frac{\gamma+1}{2\gamma}\left(v_0 + v_1\right) v,
\ee
\be\label{eq:ener_flux_pranzero}
\frac{v^2}{2} + h - \frac{\kappa}{m_0 C_p} \dv{h}{x} = \frac{\gamma+1}{\gamma - 1}\frac{v_0v_1}{2}.
\ee
Taking the spatial derivative of (\ref{eq:mom_flux_pranzero}) and eliminating the enthalpy derivative via (\ref{eq:ener_flux_pranzero}) and the enthalpy via (\ref{eq:mom_flux_pranzero}) gives
\be\label{eq:veq_pranzero}
2\left(v  - v_i\right) L_\kappa \frac{\kappa}{\kappa_0}\dv{v}{x} = \frac{\gamma + 1}{2}\left(v - v_0\right)\left(v - v_1\right).
\ee
Notice that unlike equations (\ref{eq:veq_pran34}) and (\ref{eq:veq_praninf}), equation (\ref{eq:veq_pranzero}) is singular at $v = v_i$ ($v_i$ is the isothermal sonic point for this solution). Expressed as an integral over the velocity,
\be\label{eq:integral_pranzero}
x =  \frac{4 L_\kappa}{\gamma+1} \int \frac{\left(\kappa/\kappa_0\right)\left(v  - v_i\right)}{\left(v - v_0\right)\left(v - v_1\right)}\,dv,
\ee
a solution can be obtained for $x(v)$ (again assuming constant $\kappa$):
\be\label{eq:solution_pranzero}
x =  \frac{4 L_\kappa}{\gamma\left(\gamma+1\right)} \ln \left[\left(v_0 - v\right)^{\frac{\beta v_0 - v_1}{v_0 - v_1}}\left(v - v_1\right)^{\frac{v_0 - \beta v_1}{v_0 - v_1}}\right],
\ee
where
$$
\beta \equiv \frac{3\gamma-1}{\gamma+1}.
$$
From (\ref{eq:mom_flux_pranzero}), the temperature in this limit is given by
\be\label{eq:temp_pranzero}
T = \frac{v\left(2v_i  - v\right)}{(\gamma - 1) C_v}.
\ee

\begin{figure}
\centering
\begin{tabular}{cc}
  \includegraphics[scale=0.4]{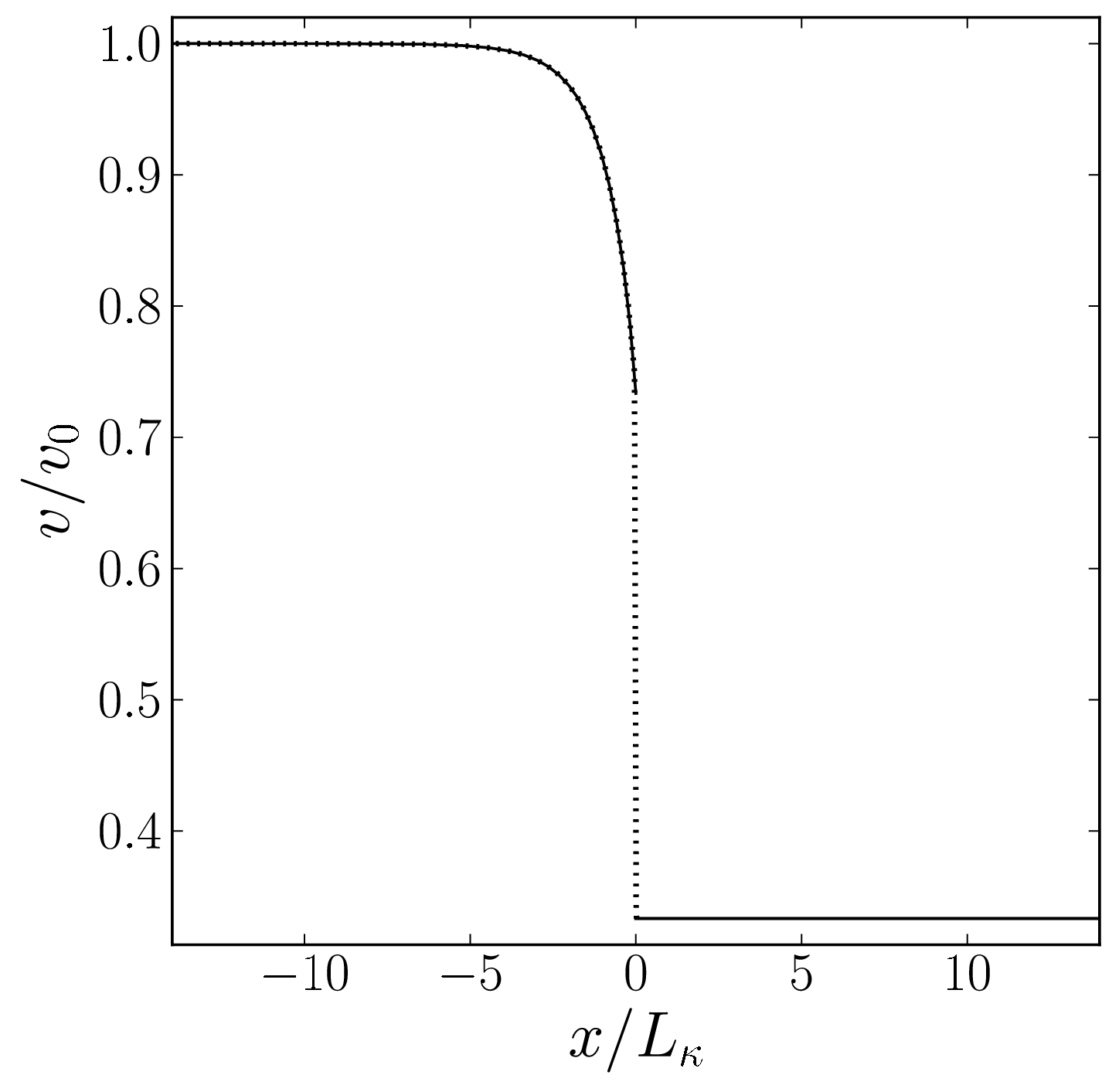} &
  \hspace{-0.0in}
  \includegraphics[scale=0.4]{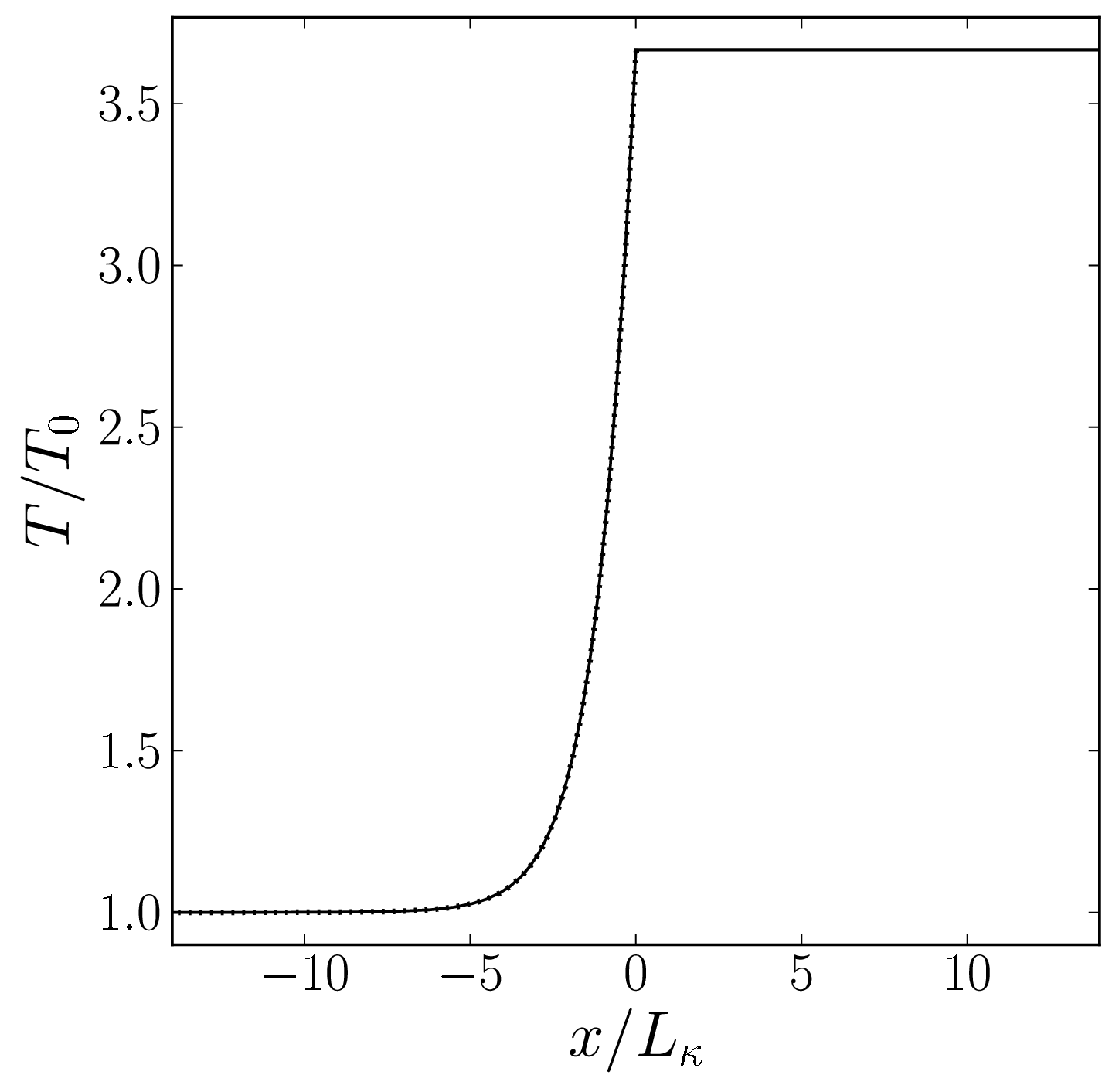}
\end{tabular}
  \caption{Velocity (\emph{left}) and temperature (\emph{right}) for the $\Pran \ttz$ solution with $M_0 = 3$ and constant conductivity. No distinction is visible between the analytical (\emph{solid}) and numerical (\emph{dotted}) results. The isothermal shock is located at $x = 0$. The material in front of the shock is heated to temperatures above the ambient temperature because heat is being conducted from the hotter post-shock region to the colder pre-shock region.}
\label{fig:solution_pranzero}
\end{figure}

As discussed in \cite{Zel'dovich02}, the small-$\Pran$ solution can be either discontinuous or continuous depending upon whether the isothermal sonic point lies inside or outside the shock region. The function $T(v)$ given by expression (\ref{eq:temp_pranzero}) passes through a maximum at $v = v_i$ and is monotonically increasing ($dT/dv > 0$) for $v_1 < v < v_i$ (see Figure 7.7 of \citealt{Zel'dovich02} for a graphical representation). The velocity in the frame of the shock (or, equivalently, the specific volume) must decrease in this region as it has not yet reached its final value, i.e., $dv/dx < 0$. This implies that the temperature also decreases in this region: $dT/dx = (dT/dv)(dv/dx) < 0$. However, this contradicts
\be
\dv{T}{x} = \frac{(\gamma + 1)\rho_0 v_0}{2 (\gamma - 1) \kappa}\left(v_0 - v\right)\left(v - v_1\right) > 0,
\ee
i.e., the temperature monotonically increases throughout the shock. The region $v_1 < v < v_i$ is thus excluded as unphysical. Since the presence of heat conduction also implies a continuous temperature, the only possibility is for the velocity to drop immediately to $v_1$ as soon as the temperature reaches $T_1$, i.e. an isothermal shock occurs. From (\ref{eq:temp_pranzero}), $T = T_1$ for $v \left(2v_i  - v\right) = v_1\left(2v_i  - v_1\right)$, or $\left(v - v_1\right)\left(v - 2v_i + v_1\right) = 0$, i.e., the embedded discontinuity occurs at
$$
v = 2v_i - v_1.
$$
If the singularity lies within the shock region, $v_i > v_1$, the small-$\Pran$ solution is given by expression (\ref{eq:solution_pranzero}) for $2v_i - v_1 < v < v_0$, followed by an isothermal shock from $v = 2v_i - v_1$ to $v = v_1$. If the singularity falls outside the shock region, $v_i < v_1$ or
\be
M_0 < \sqrt{\frac{3\gamma-1}{\gamma\left(3 - \gamma\right)}},
\ee
the solution is continuous and given by expression (\ref{eq:solution_pranzero}) throughout the shock region. Figure \ref{fig:solution_pranzero} shows the velocity and temperature for a discontinuous small-$\Pran$ solution with $M_0 = 3$ and constant conductivity. A value of $\Pran = 10^{-3}$ was used to generate the numerical results in this figure.

\subsection{Radiation heat conduction}\label{sec:grey_diffusion}

In an opaque gas, thermal radiation is in local thermodynamic equilibrium with the gas and diffuses from high to low temperature regions, thus acting as a form of heat conduction. For a constant opacity, radiation gives rise to a thermal conductivity with a $T^3$ dependence: 
$$
\kappa = \frac{16\sigma}{3\chi} T^3 ,
$$
where $\sigma$ is the Stefan-Boltzmann constant and $\chi$ is the opacity in units of inverse length \citep{Zel'dovich02}. Equation (\ref{eq:integral_pranzero}) in that case can be expressed as
$$
x =  \frac{4 L_\kappa}{\gamma+1} \int \left(\frac{T}{T_0}\right)^3\frac{v  - v_i}{\left(v - v_0\right)\left(v - v_1\right)}\,dv,
$$
with $\kappa_0 = 16\sigma T_0^3/(3\chi)$. Using equation (\ref{eq:temp_pranzero}), this can be rewritten as
\be\label{eq:x_grey}
x =  \frac{4 L_\kappa\gamma^3 M_0^6}{\gamma+1} \int \frac{\eta^3 \left(2\eta_i  - \eta\right)^3\left(\eta  - \eta_i\right)}{\left(\eta - 1\right)\left(\eta - \eta_1\right)}\,d\eta,
\ee
where $\eta \equiv v/v_0$ and $\eta_i \equiv v_i/v_0$. The integrand in (\ref{eq:x_grey}) can be expanded into
$$
\frac{\eta^3 \left(2\eta_i  - \eta\right)^3\left(\eta  - \eta_i\right)}{\left(\eta - 1\right)\left(\eta - \eta_1\right)} = \frac{\eta^6 + c_1\eta^5 + c_2\eta^3}{\eta - 1} 
+ \frac{-2\eta^6 + c_3\eta^5 + c_4\eta^3}{\eta - \eta_1} ,
$$
where
$$
c_1 \equiv \frac{7\eta_i + \eta_1 - 2 - 18 \eta_i^2}{1-\eta_1} , \;\; c_2 \equiv \frac{4\eta_i^3\left(2\eta_i - 5\right)}{\eta_1-1},
$$
$$
c_3 \equiv \frac{-7\eta_i \eta_1 - \eta_1^2 + 2\eta_1 + 18 \eta_i^2}{1-\eta_1} , \;\; c_4 \equiv \frac{4\eta_i^3\left(5 \eta_1 - 2\eta_i\right)}{\eta_1-1}.
$$
Using the result (for integer $m$)
$$
\int \frac{z^n}{z-c} dz = c^n \ln\left(z-c\right) + \sum_{m=1}^n c^{n-m}\frac{z^m}{m},
$$
the integral in (\ref{eq:x_grey}) is given by
\begin{eqnarray}\label{eq:solution_grey}
\int \frac{\eta^3\left(2\eta_i - \eta\right)^3\left(\eta  - \eta_i\right)}{\left(1-\eta\right)\left(\eta - \eta_1\right)} d\eta 
= \ln\left(1-\eta\right)^{\alpha_1} &+& \ln\left(\eta-\eta_1\right)^{-\alpha_2}
\nonumber \\
+ \sum_{m=1}^6\left(1 - 2\eta_1^{6-m}\right)\frac{\eta^m}{m} + \sum_{m=1}^5\left(c_1 + c_3\eta_1^{5-m}\right)\frac{\eta^m}{m} &+& \sum_{m=1}^3\left(c_2 + c_4\eta_1^{3-m}\right)\frac{\eta^m}{m},
\end{eqnarray}
where
$$
\alpha_1 \equiv \frac{\left(\eta_i - 1\right)\left(2\eta_i - 1\right)^3}{\eta_1-1},\;\; \alpha_2 \equiv \frac{ \eta_1^3\left(\eta_i - \eta_1\right)\left(2\eta_i - \eta_1\right)^3}{\eta_1 - 1}.
$$
Inserting this result into expression (\ref{eq:x_grey}) gives a closed form expression for $x(v)$.

\begin{figure}
\centering
\begin{tabular}{cc}
  \includegraphics[scale=0.4]{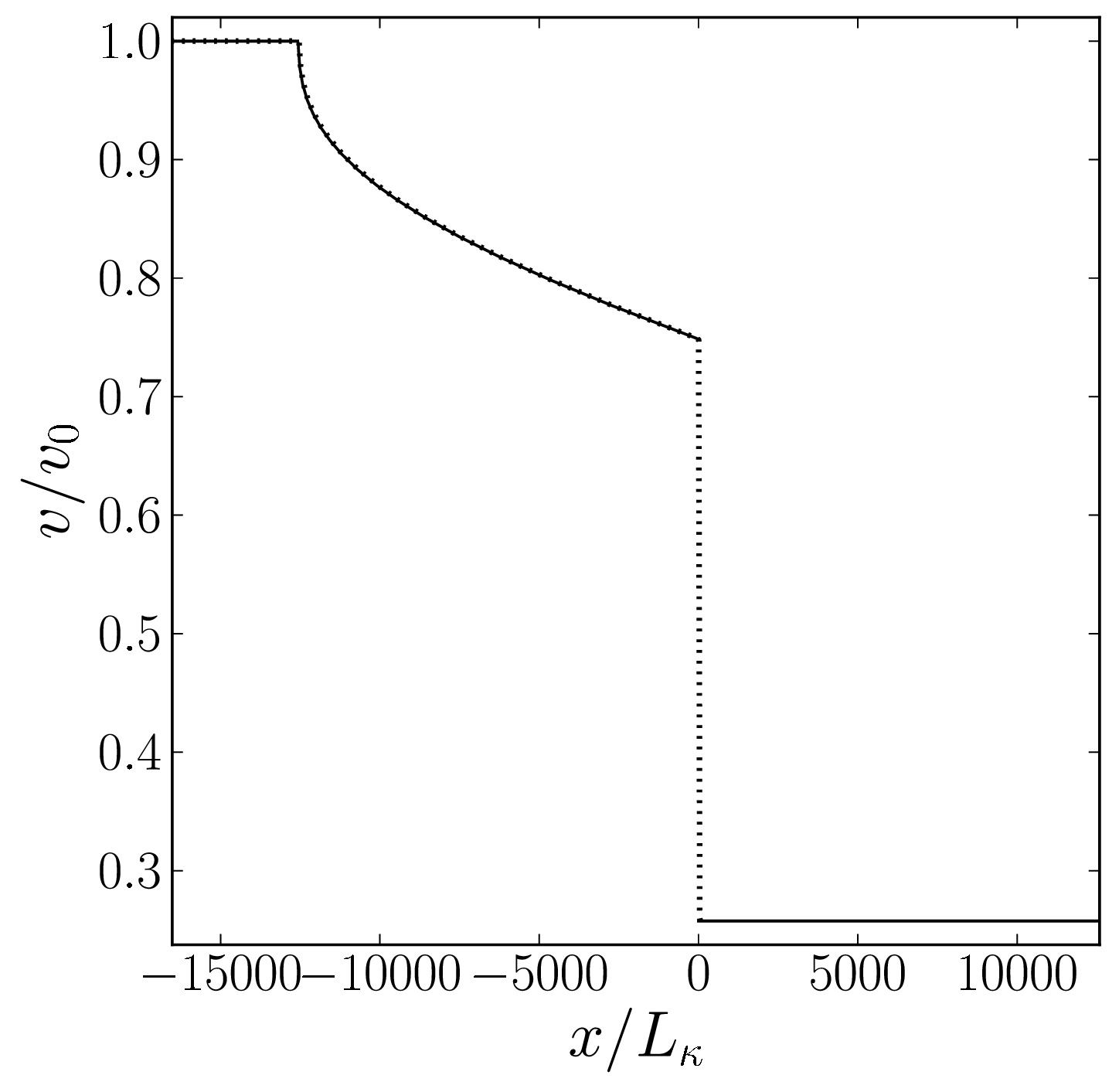} &
  \hspace{-0.0in}
  \includegraphics[scale=0.4]{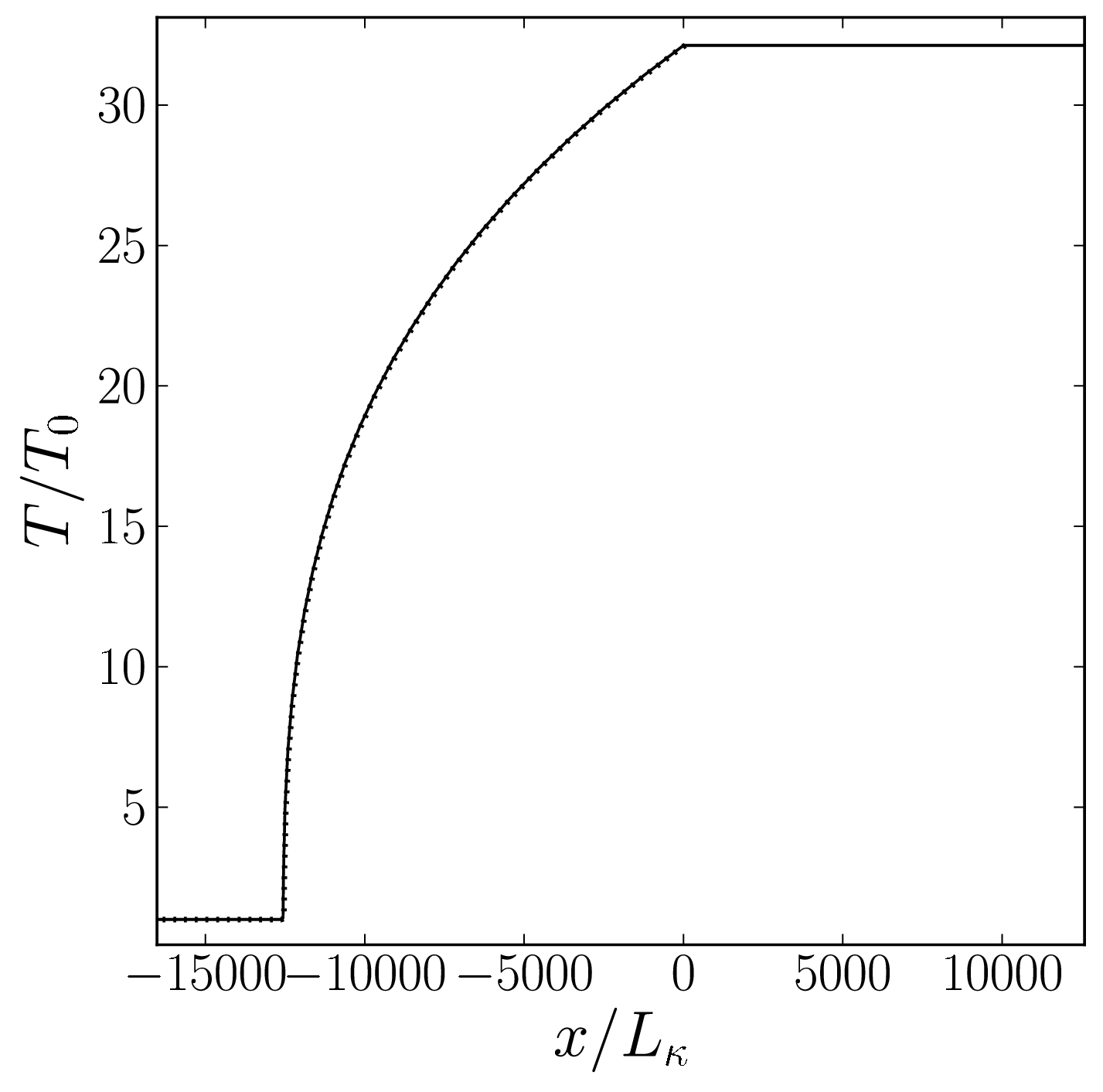}
\end{tabular}
  \caption{Velocity (\emph{left}) and temperature (\emph{right}) for the $\Pran \ttz$ solution with $M_0 = 10$ and $\kappa \sim T^3$. No distinction is visible between the analytical (\emph{solid}) and numerical (\emph{dotted}) results. The isothermal shock is located at $x = 0$. The material in front of the shock is heated to temperatures above the ambient temperature because heat is being conducted via radiation from the hotter post-shock region to the colder pre-shock region.}
\label{fig:solution_grey_M10}
\end{figure}

\begin{figure}
\centering
\begin{tabular}{cc}
  \includegraphics[scale=0.4]{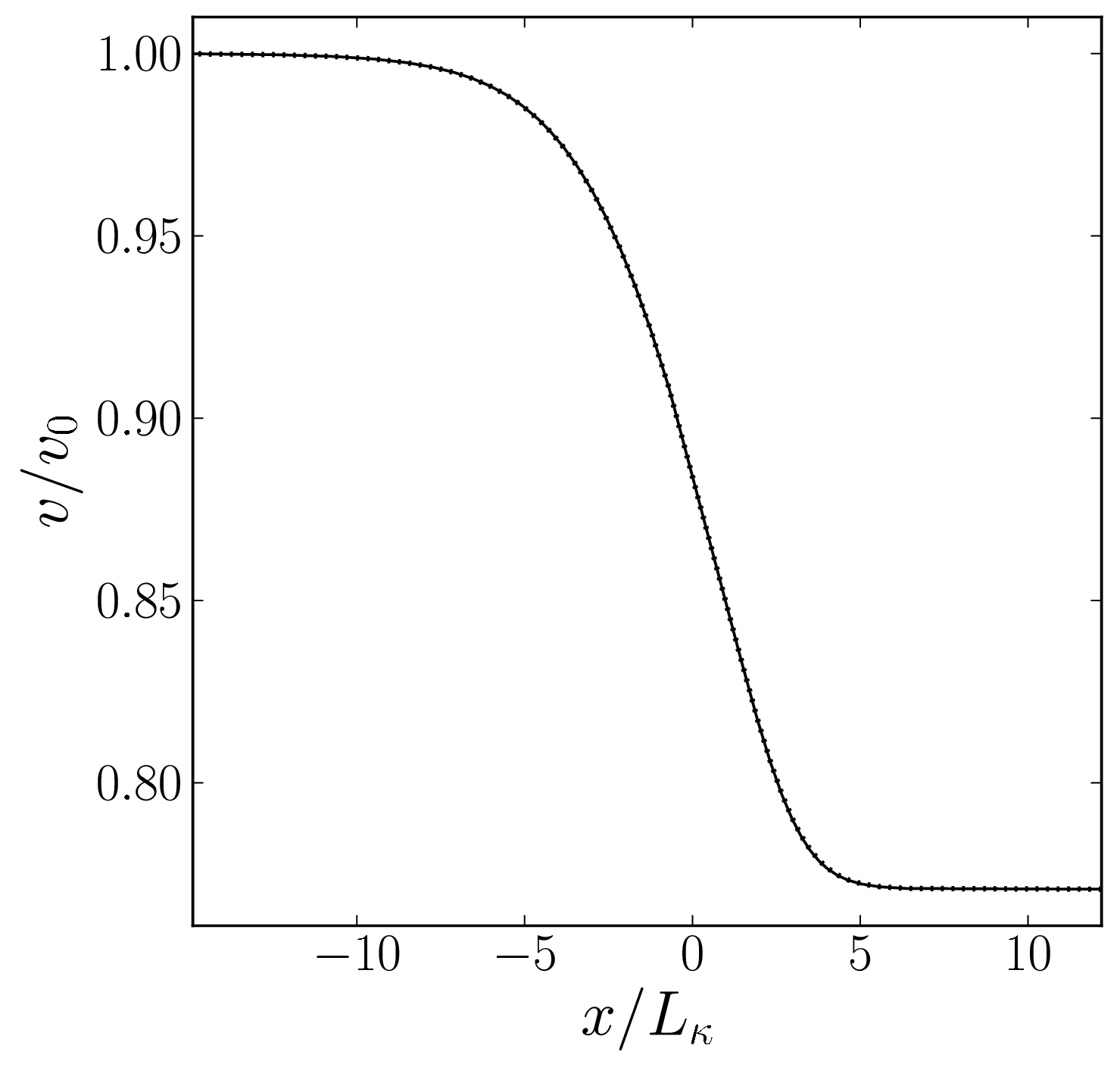} &
  \hspace{-0.0in}
  \includegraphics[scale=0.4]{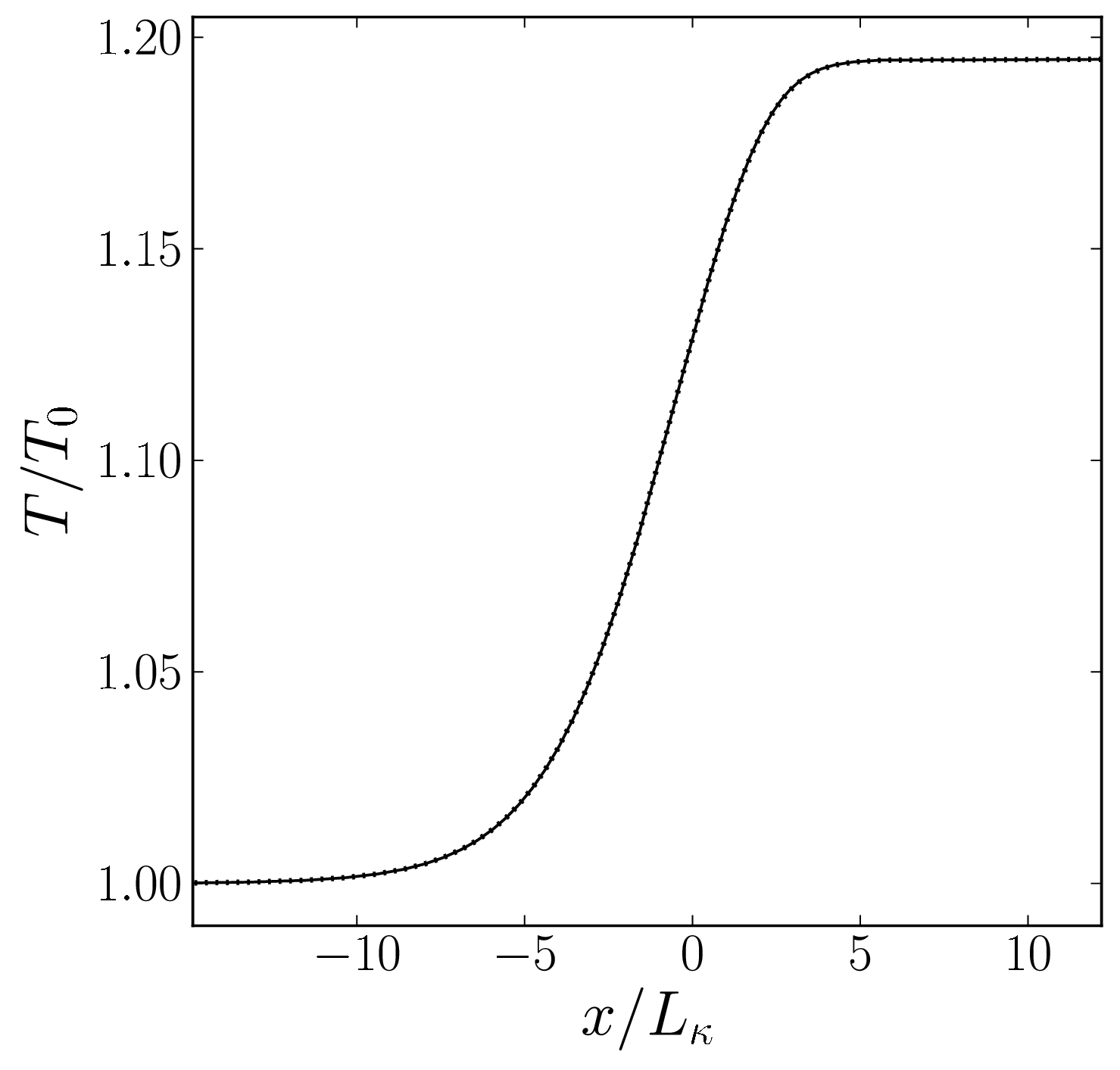}
\end{tabular}
  \caption{Velocity (\emph{left}) and temperature (\emph{right}) for the $\Pran \ttz$ solution with $M_0 = 1.2$ and $\kappa \sim T^3$. No distinction is visible between the analytical (\emph{solid}) and numerical (\emph{dotted}) results.}
\label{fig:solution_grey_M1.2}
\end{figure}

Figures \ref{fig:solution_grey_M10} and \ref{fig:solution_grey_M1.2} show the velocity and temperature for the solution described in this section with $M_0 = 10$ and $M_0 = 1.2$, respectively. A value of $\Pran = 10^{-4}$ was used to generate the numerical results in these figures. Incidentally, this is an analytical solution for radiative shocks \citep{Zel'dovich02,Lowrie07} in the limit of constant opacity and a radiation energy much lower than the gas internal energy. In the notation of \cite{Lowrie07}, this solution applies to the ${\cal P}_0 \rightarrow 0$ limit, where ${\cal P}_0$ is approximately the ratio of radiation to gas pressures. Compare figures \ref{fig:solution_grey_M10} and \ref{fig:solution_grey_M1.2} with figures 3 and 5 of \cite{Lowrie07}.

\subsection{General viscosity and conductivity}\label{sec:power_law}

Equations (\ref{eq:integral_pran34}), (\ref{eq:integral_praninf}) and (\ref{eq:integral_pranzero}) can be solved numerically for any $\kappa(\rho,T)$ and $\mu(\rho,T)$, whether analytical or tabular, using (\ref{eq:mass_flux}) and either (\ref{eq:temp_pran34}), (\ref{eq:temp_praninf}) or (\ref{eq:temp_pranzero}) to express $\rho$ and $T$ as functions of $v$. The problem can thus be reduced to quadrature under quite general conditions. For a viscosity and thermal conductivity that vary as a power-law in density and temperature,
$$
\mu = \mu_0 \left(\frac{\rho}{\rho_0}\right)^{a} \left(\frac{T}{T_0}\right)^{b},\;\; \kappa = \kappa_0 \left(\frac{\rho}{\rho_0}\right)^{a}\left(\frac{T}{T_0}\right)^{b},
$$
expressions (\ref{eq:integral_pran34}), (\ref{eq:integral_praninf}) and (\ref{eq:integral_pranzero}) become
\be\label{eq:xpl_pran34}
x\left(\Pran = 3/4\right) = \frac{2L_\kappa}{\gamma + 1}\left(\frac{\left[\gamma-1\right]M_0^2}{2}\right)^b\int \frac{\eta^{1-a}\left(R_\infty \eta_1 - \eta^2\right)^{b}}{\left(\eta - 1\right)\left(\eta - \eta_1\right)} \, d\eta,
\ee
\be\label{eq:xpl_praninf}
x\left(\Pran = \infty\right) = \frac{2L_\mu}{\gamma + 1} \left(\frac{\gamma\left[\gamma-1\right]M_0^2}{2}\right)^b\int \frac{\eta^{1-a}\left(\eta^2 - 4\eta_i\eta + R_\infty\eta_1\right)^{b}}{\left(\eta - 1\right)\left(\eta - \eta_1\right)}   \, d\eta,
\ee
\be\label{eq:xpl_pranzero}
x\left(\Pran = 0\right) = \frac{4L_\kappa}{\gamma + 1} \left(\gamma M_0^2\right)^b \int \frac{\left(\eta  - \eta_i\right) \eta^{b-a}\left(2\eta_i  - \eta\right)^{b} }{\left(\eta - 1\right)\left(\eta - \eta_1\right)}  \, d\eta.
\ee
Analytical expressions in terms of elementary functions can be obtained for particular values of $a$ and $b$ (the solution in \S\ref{sec:grey_diffusion} is an example with $a = 0$, $b = 3$), although they can be quite lengthy. The expression for a Spitzer conductivity ($a = 0$, $b = 5/2$), for example, is even longer than expression (\ref{eq:solution_grey}) and is not reproduced here \citep{Spitzer56}. The best approach for general $a$ and $b$ is to perform the quadratures in (\ref{eq:xpl_pran34})--(\ref{eq:xpl_pranzero}) numerically. Notice that $\mu$ and $\kappa$ have been assumed to have the same temperature and density dependence so that $\Pran$ is constant, for simplicity; this assumption is not necessary and is easily relaxed.

\section{Discussion}\label{sec:discussion}

\begin{figure}
\centering
\begin{tabular}{cc}
  \includegraphics[scale=0.396]{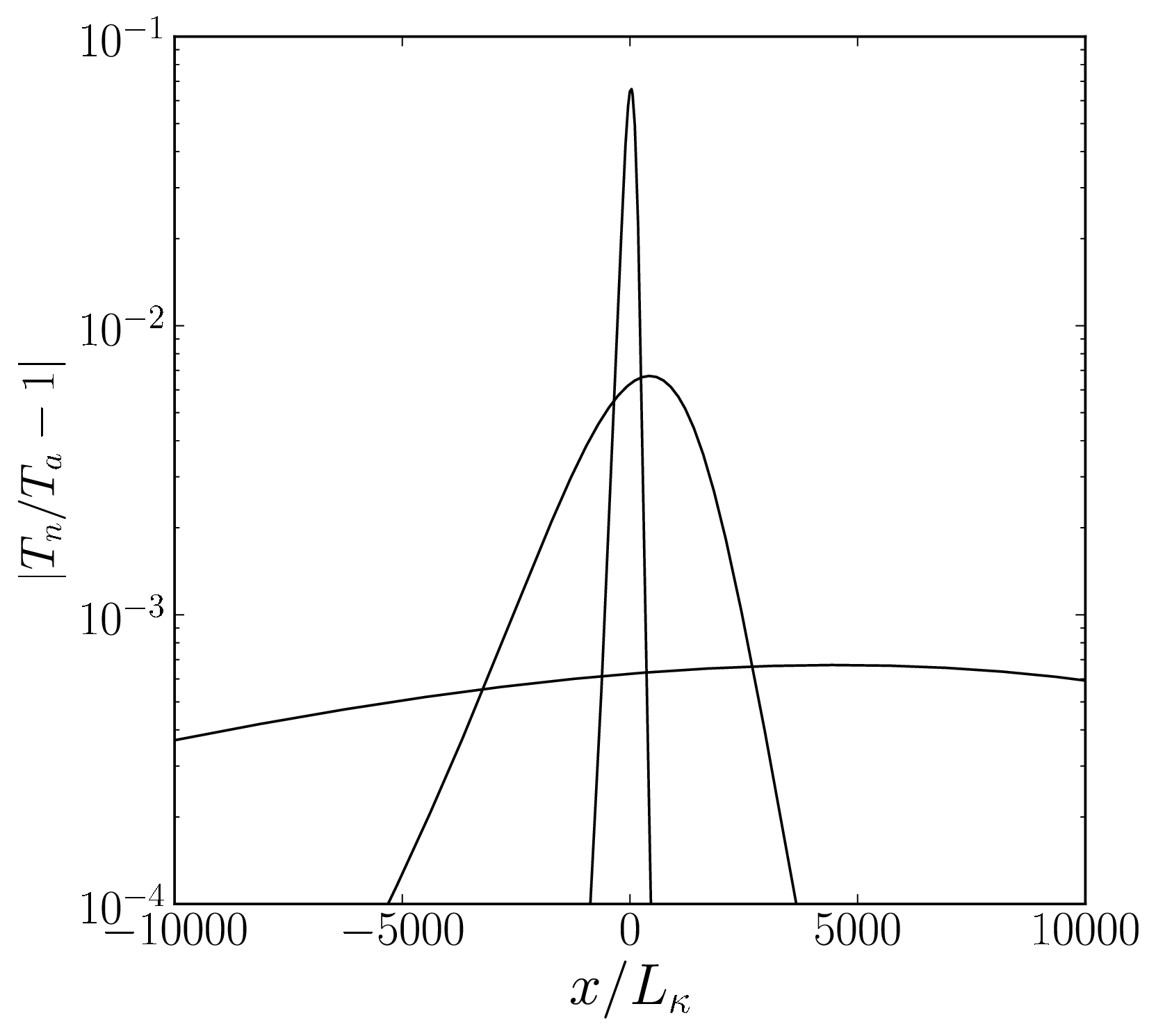} &
  \hspace{-0.0in}
  \includegraphics[scale=0.39]{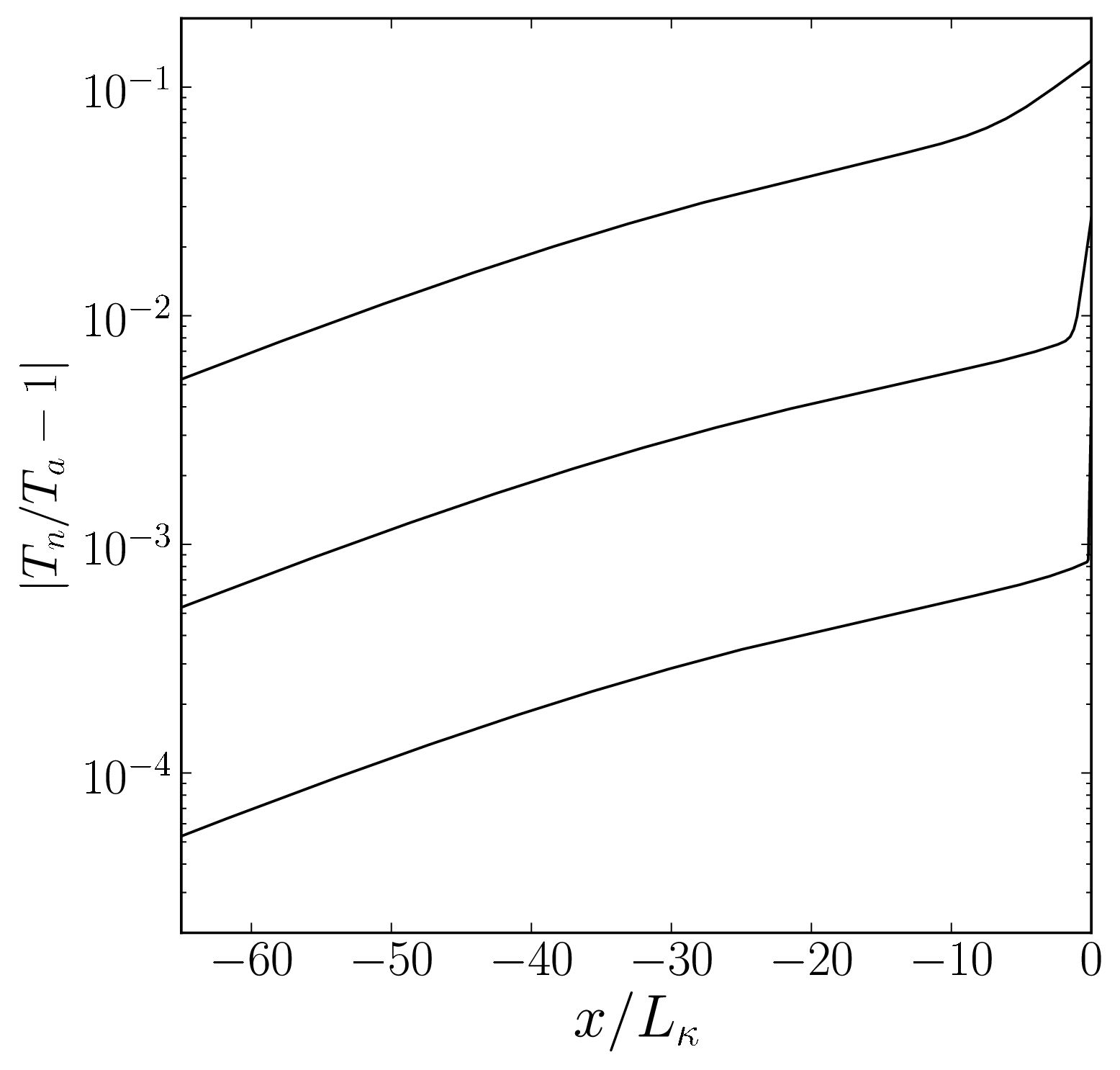}
\end{tabular}
  \caption{Temperature errors in the large-$\Pran$ (\emph{left}) and small-$\Pran$ (\emph{right}) solutions with constant viscosity and conductivity (figures \ref{fig:solution_praninf} and \ref{fig:solution_pranzero}), for (from \emph{top} to \emph{bottom}) $\Pran = 10$, $100$, $1000$ (\emph{left}) and $\Pran = 0.1$, $0.01$, $0.001$ (\emph{right}).}
\label{fig:errors}
\end{figure}

Exact solutions to the equations of fluid dynamics have been derived in the $\Pran \ttinf$ and $\Pran \ttz$ limits, analogous to the $\Pran \rightarrow 3/4$ solution derived by \cite{Becker22}. As shown in figure \ref{fig:errors}, the solutions are accurate to within \textit{O}$\left(\Pran^{-1}\right)$ for $\Pran \ttinf$ and \textit{O}$\left(\Pran\right)$ for $\Pran \ttz$. The derived solutions are given in their most general form by expressions (\ref{eq:integral_pran34}), (\ref{eq:integral_praninf}) and (\ref{eq:integral_pranzero}), along with specific forms for a constant viscosity and conductivity: (\ref{eq:solution_pran34}), (\ref{eq:solution_praninf}) and (\ref{eq:solution_pranzero}), and for a power-law temperature and density dependence: (\ref{eq:xpl_pran34})--(\ref{eq:xpl_pranzero}). The applicability of these solutions to fluids in general is limited by the use of an ideal-gas equation of state; the small-$\Pran$ solution is applicable to ideal-gas mixtures and single-component ideal gases in which temperatures are high enough for radiation heat conduction to be important. Although plasmas behave as small-$\Pran$ ideal gases, the greater mobility of the electrons relative to the ions results in separate electron and ion temperatures, a physical effect not included in this analysis \citep{Spitzer56,Zel'dovich02}. The large-$\Pran$ solution appears to be of mostly academic interest unless it can be extended to analytical equations of state appropriate for liquids and solids \citep{Ohtani09,Mozaffari13};  it remains useful, however, for code verification.

A small-$\Pran$ solution with a $T^3$ dependence has also been derived (\S\ref{sec:grey_diffusion}) that is equivalent to the semi-analytical radiative shock solutions of \cite{Lowrie07} in the limit of low radiation energy density and constant opacity. Expressions (\ref{eq:x_grey}) and (\ref{eq:xpl_pranzero}) provide a good estimate of the width of radiative shocks, the former for the constant opacity case, and the latter for a power-law opacity. In the case of a power-law opacity, simply make the substitution $a = -a^\prime-1$ and $b = 3-b^\prime$, where $a^\prime$ and $b^\prime$ are the density and temperature power-laws, respectively, for the opacity expressed in units of area per mass \citep{Bell94}. Notice that the width of a radiative shock can be quite sensitive to the shock Mach number ($x \sim M_0^6$ in the case of a constant opacity), although the applicability of the Navier-Stokes equations to large Mach number shocks is questionable \citep{Mott51,Jukes57}.

In addition to providing physical insight, the analytical solutions derived here are useful for quickly evaluating shock profiles over a wide range of parameter space. It is possible to comprehend at a glance the scaling of the solutions with various parameters without resorting to a comprehensive parameter survey via numerical integration. The solutions are also nonlinear, with the only assumptions behind their derivation being a steady-state, one planar dimension, and an ideal gas equation of state. In particular, no terms in the evolution equations have been approximated, which makes these solutions an excellent verification test for numerical algorithms.
\\

I thank the referees for their helpful comments. Many of the integrals in this work were originally obtained with Mathematica. This work was performed under the auspices of Lawrence Livermore National Security, LLC, (LLNS) under Contract No.$\;$DE-AC52-07NA27344.

\bibliographystyle{jfm}

\end{document}